\documentclass[aps,prl,twocolumn,superscriptaddress,showpacs]{revtex4}
\usepackage{amsmath}
\usepackage{amssymb}
\usepackage{graphicx}
\usepackage[dvips]{epsfig}

\newcommand{\RawBRnosyst}{\ensuremath{3.51 \pm 0.32}}

\newcommand{\RawBRfullsyst}{\pm0.29}

\newcommand{\FullBR}{\ensuremath{3.55\pm0.32{}^{\:+\:0.30}_{\:-\:{0.31}}{}^{\:+\:0.11}_{\:-\:{0.07}}}}

\newcommand{\rawYield}{\ensuremath{24100 \pm 2140\pm1260}}

\newcommand{\RawFMstat}{\ensuremath{0.026}}
\newcommand{\RawFMnosyst}{\ensuremath{2.292 \pm \RawFMstat}}

\newcommand{\FullFMsyst}{\ensuremath{0.034}}
\newcommand{\FullFM}{\ensuremath{\RawFMnosyst\pm \FullFMsyst}}

\newcommand{\RawSMstat}{\ensuremath{0.0074}}

\newcommand{\RawSMnosyst}{\ensuremath{0.0305 \pm \RawSMstat}}

\newcommand{\FullSMsyst}{\ensuremath{0.0063}}
\newcommand{\FullSM}{\ensuremath{\RawSMnosyst\pm \FullSMsyst}}

\newcommand{\GeV}{\ensuremath{\text{GeV}}}
\newcommand{\MeV}{\ensuremath{\text{MeV}}}
\newcommand{\bgs}{\ensuremath{b\to s\gamma}}
\newcommand{\ifb}{\ensuremath{\,\rm fb^{-1}}}
\begin{document}

\hbox to \textwidth{%
                 \resizebox{!}{3cm}{
                 \includegraphics{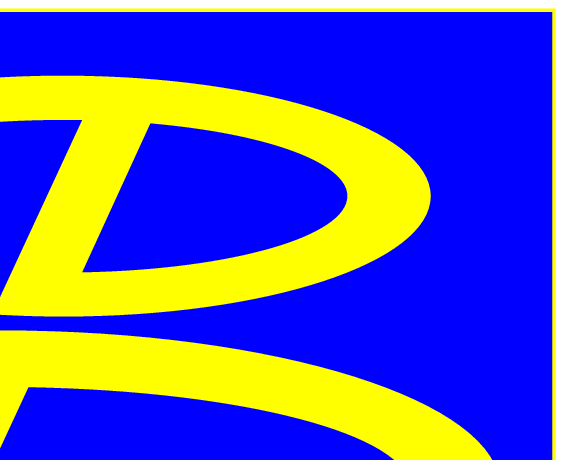}}
                 \hss{\vbox to 3cm{%
                 \hbox{Belle-Preprint-2004-6}
                 \hbox{KEK-Preprint 2003-133}
                 \hbox{hep-ex/0403004}\vss
}}}

\title{ \quad\\[1cm] \Large
           An inclusive measurement 
           of the photon
           energy spectrum in \boldmath \bgs\ decays}

\date{\today}

\affiliation{Budker Institute of Nuclear Physics, Novosibirsk}
\affiliation{Chiba University, Chiba}
\affiliation{University of Cincinnati, Cincinnati, Ohio 45221}
\affiliation{University of Frankfurt, Frankfurt}
\affiliation{University of Hawaii, Honolulu, Hawaii 96822}
\affiliation{High Energy Accelerator Research Organization (KEK), Tsukuba}
\affiliation{Hiroshima Institute of Technology, Hiroshima}
\affiliation{Institute of High Energy Physics, Chinese Academy of Sciences, Beijing}
\affiliation{Institute of High Energy Physics, Vienna}
\affiliation{Institute for Theoretical and Experimental Physics, Moscow}
\affiliation{J. Stefan Institute, Ljubljana}
\affiliation{Kanagawa University, Yokohama}
\affiliation{Korea University, Seoul}
\affiliation{Kyungpook National University, Taegu}
\affiliation{Swiss Federal Institute of Technology of Lausanne, EPFL, Lausanne}
\affiliation{University of Ljubljana, Ljubljana}
\affiliation{University of Maribor, Maribor}
\affiliation{University of Melbourne, Victoria}
\affiliation{Nagoya University, Nagoya}
\affiliation{Nara Women's University, Nara}
\affiliation{National United University, Miao Li}
\affiliation{Department of Physics, National Taiwan University, Taipei}
\affiliation{H. Niewodniczanski Institute of Nuclear Physics, Krakow}
\affiliation{Nihon Dental College, Niigata}
\affiliation{Niigata University, Niigata}
\affiliation{Osaka City University, Osaka}
\affiliation{Osaka University, Osaka}
\affiliation{Panjab University, Chandigarh}
\affiliation{Peking University, Beijing}
\affiliation{Princeton University, Princeton, New Jersey 08545}
\affiliation{RIKEN BNL Research Center, Upton, New York 11973}
\affiliation{University of Science and Technology of China, Hefei}
\affiliation{Seoul National University, Seoul}
\affiliation{Sungkyunkwan University, Suwon}
\affiliation{University of Sydney, Sydney NSW}
\affiliation{Tata Institute of Fundamental Research, Bombay}
\affiliation{Toho University, Funabashi}
\affiliation{Tohoku Gakuin University, Tagajo}
\affiliation{Tohoku University, Sendai}
\affiliation{Department of Physics, University of Tokyo, Tokyo}
\affiliation{Tokyo Institute of Technology, Tokyo}
\affiliation{Tokyo Metropolitan University, Tokyo}
\affiliation{Tokyo University of Agriculture and Technology, Tokyo}
\affiliation{University of Tsukuba, Tsukuba}
\affiliation{Utkal University, Bhubaneswer}
\affiliation{Virginia Polytechnic Institute and State University, Blacksburg, Virginia 24061}
\affiliation{Yokkaichi University, Yokkaichi}
\affiliation{Yonsei University, Seoul}
  \author{P.~Koppenburg}\affiliation{High Energy Accelerator Research Organization (KEK), Tsukuba} 
  \author{K.~Abe}\affiliation{High Energy Accelerator Research Organization (KEK), Tsukuba} 
  \author{K.~Abe}\affiliation{Tohoku Gakuin University, Tagajo} 
  \author{T.~Abe}\affiliation{High Energy Accelerator Research Organization (KEK), Tsukuba} 
  \author{I.~Adachi}\affiliation{High Energy Accelerator Research Organization (KEK), Tsukuba} 
  \author{H.~Aihara}\affiliation{Department of Physics, University of Tokyo, Tokyo} 
  \author{M.~Akatsu}\affiliation{Nagoya University, Nagoya} 
  \author{Y.~Asano}\affiliation{University of Tsukuba, Tsukuba} 
  \author{V.~Aulchenko}\affiliation{Budker Institute of Nuclear Physics, Novosibirsk} 
  \author{T.~Aushev}\affiliation{Institute for Theoretical and Experimental Physics, Moscow} 
% ...
\collaboration{The Belle Collaboration}

\begin{abstract}
We report a  fully inclusive measurement of the
flavor changing neutral current decay \bgs\ in the
energy range $1.8\,\GeV\le E^\ast_\gamma\le2.8\,\GeV$, covering
95\% of the total spectrum. Using $140\ifb$,
we obtain ${\mathcal B}(\bgs)= \left(\FullBR\right)\times 10^{-4}$,
where the errors are statistical, systematic and from theory corrections.
We also measure the first and second moments of the photon energy spectrum 
above $1.8\,\GeV$ and obtain 
$\left<E_\gamma\right>=\FullFM\,\GeV$ and 
$\left<E_\gamma^2\right>-\left<E_\gamma\right>^2=\FullSM\,\GeV^2$,
where the errors are statistical and systematic.
\end{abstract}

\maketitle



The flavor changing neutral current decay \bgs\ process is of remarkable 
theoretical interest. Its total branching fraction is very sensitive to
physics beyond the Standard Model as it may be affected by the presence of charged Higgs or SUSY particles 
in the loop. Yet the present theoretical prediction for the branching fraction
of 
  $\left(3.79{}^{\:+\:0.36}_{\:-\:0.53}\right)\times10^{-4}$~\cite{Hurth:2003dk,Gambino:2001ew}, and the average
experimental value $\left(3.3\pm0.4\right)\times10^{-4}$~\cite{PDG}
agree well. 
This agreement sets a strong constraint on, e.g., models~\cite{Isidori:2004rd} 
that accommodate the observed difference in 
CP asymmetries in the $B\to J/\psi K_S$ and $B\to \phi K_S$ 
decays~\cite{phi1}.
To obtain stronger constraints on physics beyond the
Standard Model, more precise theoretical predictions and experimental 
measurements are needed. 

On the other hand, the photon energy spectrum is almost insensitive to 
physics beyond the Standard Model~\cite{Kagan}.
At the parton level, the photon is monochromatic with energy 
$E\approx m_b/2$ in the $b$ quark rest frame. 
The energy is smeared by the motion of the $b$
quark inside the $B$ meson and gluon emission.
A measurement of the moments of this spectrum allows a 
determination of the $b$-quark mass and of its motion. 
This information can then be used to extract the CKM matrix elements 
$|V_{cb}|$ and $|V_{ub}|$ from inclusive 
semileptonic $b$ decays~\cite{NoteOn,Vcb}. 
However, a measurement of the low-energy tail of the photon
spectrum is important in this context~\cite{Bigi:2002qq}.

Belle has previously measured the \bgs\ branching fraction
with $5.8\ifb$ of data using a semi-inclusive approach. 
Because we applied an effective cut $E_\gamma>2.24\,\GeV$ 
in the $B$ rest frame, the precision of that measurement is 
limited by theoretical errors due to the extrapolation to the whole 
energy spectrum. More recently the CLEO collaboration has reported
a measurement of the branching fraction and the 
energy spectrum moments performed in a fully inclusive way~\cite{CLEOb2g} 
for the range $E_\gamma^\ast>2.0\,\GeV$ in the center-of-mass frame~\cite{CM}.
Here we present a measurement using a similar approach, but based on a much larger dataset
allowing a detailed study of the backgrounds. We extend the photon energy range 
to $E_\gamma^\ast>1.8\,\GeV$, covering almost the entire spectrum.

The \bgs\ decay is studied using the Belle detector at the KEKB
asymmetric $e^+e^-$ storage ring~\cite{KEKB}. The data consists of
a sample of $140\ifb$ taken at the $\Upsilon(4S)$ resonance
corresponding to $(152.0{}^{\:+\:0.6}_{\:-\:0.7})\times10^6$ $B\bar{B}$ pairs. Another
$15\ifb$ sample has been taken at an energy $60\,\MeV$ below the resonance 
and is used to measure the non-$B\bar{B}$ background.
Throughout this Letter, we refer to these data samples as the ON and OFF
samples, respectively.

%
The Belle detector is a large-solid-angle magnetic spectrometer
described in detail elsewhere~\cite{Belle}.
The main component relevant for this analysis is the 
electromagnetic calorimeter (ECL) made of 
$16.2$ radiation lengths long CsI(Tl) crystals. The photon energy resolution
is about 2\% for the energy range relevant in this analysis.


The strategy to extract the signal \bgs\ spectrum is to collect all 
high-energy photons, vetoing those originating from $\pi^0$ and $\eta$ 
decays to two photons. The contribution from continuum $e^+e^-\to q\bar{q}$
($q=u,d,s,c$) events is subtracted using the OFF sample. The remaining 
backgrounds from $B\bar{B}$ events are subtracted using 
Monte-Carlo (MC) distributions scaled by data control samples.


Photon candidates are selected from ECL clusters of $5\times5$ crystals
in the barrel region ($-0.5\le\cos\theta\le0.84$, where $\theta$ 
is the polar angle with respect to the beam axis). 
They are required to have an energy 
$E^\ast_\gamma$ larger than $1.5\,\GeV$, and $95$\% of the energy has to
be deposited in the central $3\times3$ crystal array.
We require isolation cuts 
to veto photons from bremsstrahlung and interaction with matter.
The center of the cluster has to be displaced
from any other ECL cluster with $E>20\,\MeV$ by at least $30\,\rm cm$ 
at the surface of the calorimeter, 
and from any reconstructed track by $3\,\rm cm$,
or by $50\,\rm cm$ for tracks with a measured momentum above $1\,\GeV/c$. 
Moreover, the angle between the photon and the highest energy lepton
in the event has to be larger than $0.3$ radians at the interaction point.
We veto candidate photons from $\pi^0$ and $\eta$ decays to two photons by
combining them with any other photon. We reject the pair if the likelihood
of being a $\pi^0$ or $\eta$ is larger than $0.1$ and $0.2$, respectively.
These likelihoods are determined from MC and are functions of the 
laboratory energy of the other
photon, its polar angle $\theta$ and the mass of the two-photon system.

In order to reduce the contribution from continuum events,
we use two Fisher discriminants. 
The first exploits the spherical shape of $B\bar{B}$ events and 
is built using ten event-shape variables. These
variables are calculated using either all tracks and showers
in the event or excluding the photon candidate. The event
shape variables include Fox-Wolfram moments, thrust, and the angles
of the thrust axis with respect to the beam and photon direction.
The second discriminant exploits the topology of 
\bgs\ events and combines three energy flows around the photon axis.
These energy flow variables are obtained using all particles, except 
for the photon candidate, whose direction lies in the three regions
defined by $\alpha^\ast<30^\circ$, $30^\circ\le\alpha^\ast\le140^\circ$,
$\alpha^\ast>140^\circ$, where $\alpha^\ast$ 
is the angle to the candidate photon.

To optimize these selection criteria, we use a Monte Carlo
simulation~\cite{MC} containing large samples of $B\bar{B}$, $q\bar{q}$ and
signal weighted according to the luminosities of the ON and OFF
samples.
The signal MC is generated as a weighted sum
of $B\to K^\ast\gamma$ decays, where $K^\ast$ is any known spin-$1$ resonance
with strangeness $S=1$. The relative weights are obtained by fitting 
the total photon spectrum to a theoretical model~\cite{Kagan}.
The signal MC is normalized to the average measured
branching fraction~\cite{PDG}. 
To improve the understanding of the photon energy spectrum 
at low energies, the selection criteria are
optimized to maximize the sensitivity to the signal in
the energy bin $1.8\,\GeV<E^\ast_\gamma<1.9\,\GeV$.

After these selection criteria we observe $1.2\times10^6$ 
photon candidates in ON and $1.1\times10^5$ in OFF data. 
The spectrum measured in OFF data is scaled by luminosity to 
the expected number of non-$B\bar B$ events in ON data and subtracted. 
To take into account the effect of the $60\,\MeV$ ($0.5$\%) energy
difference, the measured OFF energies are scaled by 
an empirical factor of $1.004$ obtained from a MC study.
The ON and scaled OFF spectra and their difference are shown in 
Fig.~\ref{fig:Four}.
\begin{figure}[!t]
\centerline{
\epsfxsize 3.0 truein \epsfbox{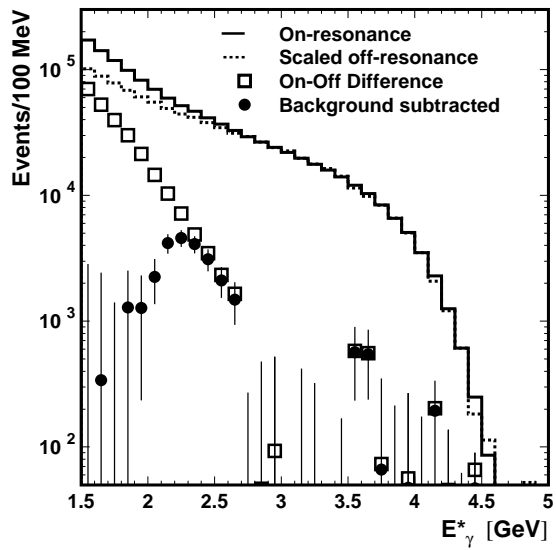}
}   
\caption{Photon energy spectra in the $\Upsilon(4S)$ frame.}
\label{fig:Four}
\end{figure}

We then subtract the backgrounds from $B$ decays from the obtained spectrum.
Five background categories are considered: 
{\it (i)} photons from $\pi^0\to\gamma\gamma$, which account for more than
half of the background in the $1.8$--$2.8\,\GeV$ range;
{\it (ii)} photons from $\eta\to\gamma\gamma$;
{\it (iii)} other real photons (mainly decays of $\omega$, $\eta'$, and $J/\psi$, 
and bremsstrahlung);
{\it (iv)} ECL clusters not due to single photons (mainly electrons 
interacting with matter, $K^0_L$ and $\bar{n}$);
{\it (v)} beam background.
For each of these categories we take the predicted background
from MC and scale it according to measured yields wherever possible.
The inclusive $B\to\pi^0X$ and $B\to\eta X$ spectra are measured in data
using pairs of photons with a well-balanced energy
and applying the same ON$-$OFF subtraction procedure. 
The yields obtained are $5$ to $15$\% larger 
than MC expectations depending on the photon energy range.
Since there is good agreement between MC and data for all features
of the GEANT simulation for photons and electrons,
we believe that the observed discrepancy between the measured and
simulated $\pi^0$ spectrum is due to the generator~\cite{Pi0spectrum}.
Beam background is measured using a sample of randomly triggered 
events and added to the $B\bar{B}$ MC. 

For each selection criterion and each background
category we determine the $E^\ast_\gamma$-dependent selection efficiency 
in OFF-subtracted ON data and MC using appropriate control samples. We then
scale the MC background sample according to the ratio
of these efficiencies. 
The efficiencies of the $\pi^0$ and $\eta$ vetoes for non-$\pi^0$, non-$\eta$ photons 
          are measured in data using one photon from a well reconstructed
          $\pi^0$ applying the veto without using the other photon of the pair.
The $\pi^0$ veto efficiency is measured
          using a sample of photons coming from measured $\pi^0$ decays.
	  We use partially reconstructed $D^{\ast+}\to D^0\pi^+$, $D^0\to K^-\pi^+\pi^0$
          decays where the $\pi^0$ is replaced by the candidate photon
	  in the reconstruction. 
The $\eta$ veto efficiency for photons from $\pi^0$'s and event-shape criteria efficiencies 
          are measured using a $\pi^0$ anti-veto sample. It is made of photons passing
	  all selection criteria except the $\pi^0$ veto,
          which are combined with another photon in the event to give 
	  a $\pi^0$-likelihood larger than $0.75$.
Other efficiencies are measured using the signal sample. 

The ratios of data and MC efficiencies versus $E^\ast_\gamma$ are fitted using first or
second order polynomials, which are used to scale the background MC. 
Most are found to be statistically compatible with unity.
An exception is the efficiency of the requirement that $95$\% of the energy
be deposited in the central nine cells of the $5\times5$ cluster,
which is found to be poorly modelled by our MC for non-photon
backgrounds. 
We estimate the efficiency for data using a sample 
of candidate photons in OFF-subtracted ON data after subtracting the
known contribution from real photons.
This 
increases the
yield of background {\it (iv)} by 50\%.
The yield from the five background categories, after having been 
properly scaled by the above described procedures, are subtracted from 
the OFF-subtracted spectrum. The result is shown in Fig.~\ref{fig:Four}. 

The spectrum contains \rawYield\ events in the $1.8$--$2.8\,\GeV$ 
energy range, where the two errors are the statistical error of the 
OFF-subtracted ON data and of the $B\bar{B}$ background subtractions, and the systematic error 
related to the data/MC efficiency ratio fits used in 
the $B\bar{B}$ background scaling. 
We correct this spectrum for the signal selection efficiency function
obtained from signal MC, applying the same data/MC correction factors
as for the generic photon background category {\it (iii)}.
The average signal selection efficiency is $23$\%.

\begin{figure}[!b]
\centerline{
\epsfxsize 3.0 truein \epsfbox{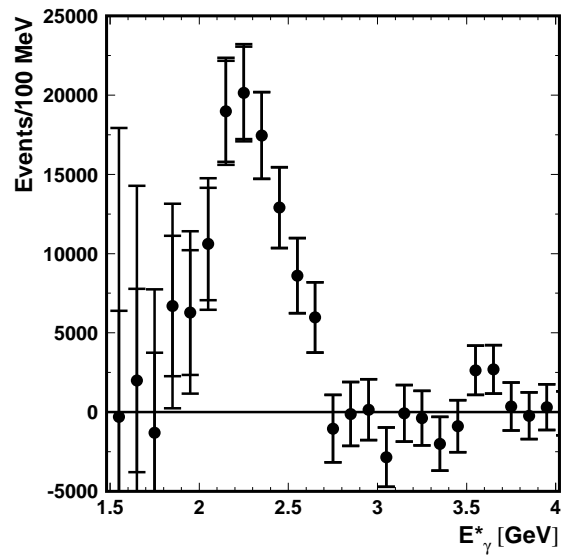}
}   
\caption{Efficiency-corrected photon energy spectrum.
         The two error bars show the statistical and total errors.}
\label{fig:figure2}
\end{figure}

The efficiency-corrected spectrum is shown in Figure~\ref{fig:figure2}. 
The two error bars for each point show the statistical
and the total error, including the systematic error which is correlated among the points.
As expected, the spectrum above the endpoint for decays of 
$B$ mesons from the $\Upsilon(4S)$ at about $3\,\GeV$, 
is consistent with zero.
Integrating this spectrum from $1.8$ to $2.8\,\GeV$, we obtain a
partial branching fraction of 
$\left(\RawBRnosyst\RawBRfullsyst\right)\times10^{-4}$.

\begin{table}[t]
  \begin{tabular}{| l | r |}
     \hline
     Source of systematic error & $\times 10^{-4}$  \\
     \hline\hline
     Raw branching fraction             & $\RawBRnosyst$  \\
     \hline
     Data/MC efficiency ratio fits      & $\pm\:0.208$    \\                     
     Choice of fitting functions        & $\pm\: 0.048$ \\ 
     Number of $B\bar{B}$-events  & ${}^{\:+\:0.139}_{\:-\:{0.160}}$ \\  
     ON-OFF data subtraction            & $\pm\: 0.026$ \\ 
     Other $B\bar{B}$ photons           & $\pm\: 0.054$ \\ 
     $\eta$ veto efficiency on $\eta$   & $\pm\: 0.008$ \\ 
     Signal MC                          & $\pm\: 0.089$ \\ 
     Photon detection efficiency        & $\pm\: 0.072$ \\ 
     Energy leakage                     & ${}^{\:+\:0.035}_{\:-\:{0.000}}$ \\ 
     \hline
     Total error for partial ${\mathcal B(b\to q\gamma)}$ & ${}^{\:+\:0.282}_{\:-\:{0.291}}$ \\
     \hline
  \end{tabular}
  \caption{Overview of systematic errors.
           \label{Tab:systematics}}
\end{table}
The sources of systematic error are listed in Table~\ref{Tab:systematics}.
They are added in quadrature.
The largest sources are the errors of the data/MC efficiency ratio fits
($5.9$\% of the signal yield).
For the error related to the choice of the polynomial 
    functions in the data/MC efficiency ratio fits, 
    we perform the same fit increasing the polynomial order by one.
The number of $B\bar{B}$ events is determined from the number
of hadronic events in ON and OFF data. The
relative luminosities of the two samples are determined from
radiative Bhabhas and $e^+e^-\to\mu^+\mu^-$ events.
The errors on the OFF data subtraction are estimated using the 
    result of the fit to the spectrum above the endpoint.
    We integrate the resulting function in the $1.8$--$2.8\,\GeV$ range
    and obtain a yield of $+40\pm160$. We add $\pm200$ ($0.8$\%)
    to the systematic error. 
As we do not measure the yields of photons from sources other than 
    $\pi^0$'s and $\eta$'s in $B\bar{B}$ events, we vary the 
    expected yields of these additional sources by $\pm20$\%.
For the model dependence of signal selection efficiency we
    use an alternate signal MC that favors high-mass resonances
    decaying into high-multiplicity final states.
    Using this MC to correct for the efficiency changes the branching
    fraction by $\pm2.5$\%. 
The error on the photon detection efficiency in the ECL is
    measured to be $2.3$\% using radiative Bhabha events. 
    This error also affects the estimation of photons from
    $B\bar{B}$ and contributes $\pm2.0$\% to the
    systematic error.
Due to the low-energy tail in the photon energy measurement,
    some part of the spectrum 
    may lie below the range of integration.
    We estimate this fraction to be 
    smaller than $1$\%.
    As this value is shape-dependent, we 
    do not correct the measured branching fraction for it but instead add
    a $({}^{+1.0}_{-0.0})$\% systematic error.

In order to obtain the total \bgs\ branching fraction we apply
corrections for the contribution from Cabibbo suppressed $b\to d\gamma$
decays and for the invisible part of the spectrum 
below $1.8\,\GeV$.
The ratio of the \bgs\ and $b\to d\gamma$ branching fractions 
is assumed to be 
$R_{d/s}=(3.8\pm0.6)$\%~\cite{Hurth:2003dk}.
%
The selection efficiency for $b\to d\gamma$ is found to be equal
to the efficiency for \bgs\ within 10\%, which we include in the
systematic error.
The fraction of the spectrum above $1.8\,\GeV$ is assumed
to be $R_{1.8}=0.952{}^{\:+\:0.013}_{\:-\:0.029}$ from Gambino and 
Misiak~\cite{Gambino:2001ew}. As a cross-check we also use the value from Kagan and 
Neubert~\cite{Kagan}
$R_{1.8}=0.958{}^{\:+\:0.013}_{\:-\:0.029}$ , 
and $R_{1.8}=0.95\pm0.01$ from Bigi and Uraltsev~\cite{Bigi:2002qq}.
We combine the errors on $R_{d/s}$, $R_{1.8}$ and the 
difference between the  $R_{1.8}$ values into the theoretical error.
With these two corrections, we obtain
$$
  {\mathcal B}\left(\bgs\right) = \left(\FullBR\right)\times 10^{-4}
$$
for the total \bgs\ branching fraction.
This result is in good agreement with theoretical expectations
and with previous experimental measurements~\cite{Belleb2g,CLEOb2g}.

We also measure the first two moments of the energy spectrum
in the $B$ rest frame.
We extract the raw moments from the distribution shown
in Fig.~\ref{fig:figure2} in the range $1.8\,\GeV\le E^\ast_\gamma\le 2.8\,\GeV$
and correct them for the effect of the boost of the $B$ meson in the
$\Upsilon(4S)$ frame, for the energy resolution and for the 
$100\,\MeV$ binning. We do not correct the moments for the missing
low-energy tail. We obtain the following moments for 
$E^\ast_\gamma>1.8\,\GeV$ (corresponding to
$E_\gamma>1.815\,\GeV$ in the $B$ rest frame):
\begin{eqnarray*}
  \left< E_\gamma \right> & = &  \FullFM\,\GeV \\
   \left<E_\gamma^2\right>-\left<E_\gamma\right>^2 & = & \FullSM \,\GeV^2,
\end{eqnarray*}
where the errors are statistical and systematic.

The systematic error contains the errors related to the moments
corrections and the error sources already mentioned  
for the branching fraction extraction. For the first moment
the systematic error is dominated by the
data/MC efficiency ratio fits ($\pm0.9$\%) and the shape of the
energy resolution ($\pm1.0$\%). The error on the second moment 
is dominated by the data/MC efficiency ratio fits ($\pm17$\%).
These results agree within $1\sigma$ with the 
only previous measurement, done by the CLEO collaboration~\cite{CLEOb2g}.
However, it should be noted that the CLEO results are 
obtained for $E^\ast_\gamma > 2.0\,\GeV$.

In conclusion, we have measured the branching fraction 
and photon energy spectrum of \bgs\ in the 
energy range $1.8\,\GeV\le E^\ast_\gamma\le2.8\,\GeV$ in a fully 
inclusive way. For the first time $95$\% or more of the spectrum
is measured, allowing the theoretical uncertainties to be
reduced to a very low level. 
Using $140\ifb$ of data taken at the $\Upsilon(4S)$ and
$15\ifb$ taken below the resonance, we obtain 
${\mathcal B}\left(\bgs\right) = \left(\FullBR\right)\times 10^{-4}$,
where the errors are statistical, systematic and theoretical, respectively. 
This result is in good agreement with the latest theoretical 
calculations~\cite{Hurth:2003dk,Gambino:2001ew}.
We have also measured the moments of the distribution and obtain
$\left< E_\gamma \right>  =   \FullFM\,\GeV$ and 
$\left<E_\gamma^2\right>-\left<E_\gamma\right>^2 = \FullSM \,\GeV^2$ 
for $E^\ast_\gamma>1.8\,\GeV$,
where the errors are statistical and systematic. 

We thank
T.~Hurth,
I.~Bigi, 
A.~Kagan and 
M.~Misiak 
for helpful discussions and correspondences. 
   We thank the KEKB group for the excellent
   operation of the accelerator, the KEK Cryogenics
   group for the efficient operation of the solenoid,
   and the KEK computer group and the NII for valuable computing and
   Super-SINET network support.  We acknowledge support from
   MEXT and JSPS (Japan); ARC and DEST (Australia); NSFC (contract
   No.~10175071, China); DST (India); the BK21 program of MOEHRD and the
   CHEP SRC program of KOSEF (Korea); KBN (contract No.~2P03B 01324,
   Poland); MIST (Russia); MESS (Slovenia); NSC and MOE (Taiwan); and DOE
   (USA).


\begin{thebibliography}{99}
\bibitem{CM}
Throughout this Letter, variables with an asterisk are 
calculated in the center of mass frame.
\end{thebibliography}
\end{document}